\documentclass[useAMS,usenatbib]{mn2e}
\usepackage{subfigure}
\usepackage{graphicx}

\title[SED sensitivity to clumpy torus properties]{Investigating the sensitivity of observed spectral energy 
distributions to clumpy torus properties in Seyfert galaxies}
\author[C. Ramos Almeida et al.]
{\parbox{\textwidth}{C. Ramos Almeida$^{1,2}$\thanks{Marie Curie Fellow. E-mail: cra@iac.es},
A. Alonso-Herrero$^{3,4}$,
N. A. Levenson$^{5}$,
A. Asensio Ramos$^{1,2}$,
J. M. Rodr\' iguez Espinosa$^{1,2}$,
O. Gonz\'alez-Mart\' in$^{1,2}$,
C. Packham$^6$,
M. Mart\' inez$^7$
}\vspace{0.4cm}\\
\parbox{\textwidth}{$^{1}$Instituto de Astrof\' isica de Canarias, Calle V\' ia L\'actea, s/n, E-38205, La Laguna, Tenerife, Spain\\
$^{2}$Departamento de Astrof\' isica, Universidad de La Laguna, E-38206, La Laguna, Tenerife, Spain\\
$^{3}$Instituto de F\' isica de Cantabria, CSIC-Universidad de Cantabria, E-39005, Santander, Spain\\
$^{4}$Augusto Gonz\'alez Linares Senior Research Fellow\\ 
$^{5}$Gemini Observatory, Casilla 603, La Serena, Chile\\ 
$^{6}$Department of Physics and Astronomy, University of Texas at San Antonio, One UTSA Circle, San Antonio, USA \\
$^{7}$Instituto Nacional de Astrof\' isica, \'Optica y Electr\'onica (INAOE), 72000 Puebla, M\'exico 
}}

\begin{document}

\date{}

\pagerange{\pageref{firstpage}--\pageref{lastpage}} \pubyear{2013}

\maketitle

\label{firstpage}

\begin{abstract}
We present nuclear spectral energy distributions (SEDs) from 1 to 18 \micron~of a small sample of nearby, nearly face-on 
and undisturbed Seyfert galaxies without prominent nuclear dust lanes. These nuclear SEDs 
probe the central $\sim$35 pc of the galaxies, on average, and include photometric and spectroscopic infrared (IR) data. 
We use these SEDs, the clumpy torus models of Nenkova et al. and a Bayesian approach to study the sensitivity of 
different IR wavelengths to the torus parameters. We find that high angular resolution 8--13 \micron~spectroscopy alone 
reliably constrains the number of clumps and their optical depth (N$_0$ and $\tau_V$). On the other hand, we need a 
combination of mid- and near-IR subarcsecond resolution photometry to
constrain torus width and inclination, as well as the radial distribution of the clouds ($\sigma$, $i$ and $q$). For 
flat radial profiles ($q$=0, 1), it is possible to constrain the extent of the mid-IR-emitting dust within the torus 
($Y$) when N-band spectroscopy is available, in addition to near-IR photometry. 
Finally, by fitting different combinations of average and individual Seyfert 1 and Seyfert 2 data, 
we find that, in general, for undisturbed, nearly face-on Seyferts without prominent nuclear dust lanes, the minimum combination 
of data necessary to reliably constrain all the torus parameters is J+K+M-band photometry + N-band spectroscopy.
\end{abstract}

\begin{keywords}
galaxies: active -- galaxies: nuclei -- galaxies: Seyfert -- galaxies: individual (NGC\,1068, NGC\,1365, 
NGC\,3081, NGC\,3227, NGC\,4151, NGC\,5643).
\end{keywords}

\section{Introduction}
\label{intro}

Recent success in explaining several properties of the nuclear infrared (IR) spectral energy distributions (SEDs) of 
Seyfert galaxies has been gathered with the assumption of a clumpy distribution of dust surrounding active galactic nuclei (AGN).
Clumpy torus models \citep{Nenkova02,Nenkova08a,Nenkova08b,Honig06,Honig10b,Schartmann08,Schartmann09,Stalevski12}
propose that the dust is distributed in clumps, instead of homogeneously filling the torus volume. 
The latter authors explain in detail how different model parameters control the shape of the 
synthetic SEDs, and in some cases compare 
with observations. 

From the observers point of view, our aim is to reproduce nuclear SEDs of nearby AGN
with torus models. In turn the goal is to look for trends in the torus model parameters with different observational AGN properties 
such as luminosity, AGN class and presence or not of broad lines
(\citealt{Mason06,Mason09,Mason13,Mor09,Horst08,Horst09,Nikutta09}, Ramos Almeida et al. 2009, 2011a,b; 
\citealt{Honig10b}; Alonso-Herrero et al.~2011, 2012, 2013; \citealt{Lira13}).

In a series of papers, we performed detailed fits to the nuclear near- and mid-IR (NIR and MIR) 
emission of Seyfert galaxies using the clumpy torus models of \citet{Nenkova08a,Nenkova08b} and a 
Bayesian approach (see below). In the first two 
works of the series (Ramos Almeida et al. 2009, 2011a; hereafter \citealt{Ramos09} and \citealt{Ramos11} respectively), we only fitted 
subarcsecond resolution photometry from 1--20 \micron. The main results of these studies are: 1) the NIR/MIR emission of 
Seyfert galaxies can be reproduced with small torus sizes, ranging from 1 to 6 pc. 2) Seyfert 2 (Sy2) tori generally have 
larger covering factors and smaller escape probabilities than those of Seyfert 1 (Sy1) galaxies. 
3) The classification of a Seyfert galaxy as a Type-1 or Type-2 might 
depend, not only on torus inclination, but also on the intrinsic properties or that torus. 

In the third paper of the series (Alonso-Herrero et al.~2011; hereafter \citealt{Alonso11}), we combined ground-based 
subarcsecond MIR spectroscopy and NIR/MIR photometry of 13 Seyfert galaxies at a median distance of 31 Mpc. 
In general, the photometry+spectroscopy fits resulted in more constrained torus widths ($\sigma$), 
torus radial extents ($Y$) and viewing angles ($i$). From the study presented in \citealt{Alonso11}, we
learnt that clumpy torus models provide good fits only for those Seyferts with low--to--moderate 
amounts of foreground extinction (A$_V\la$ 5--10 mag). The 8--13 \micron~spectra of galaxies with very deep silicate features, 
normally hosted in highly inclined or interacting galaxies and/or showing prominent dust lanes, 
cannot be reliably reproduced with clumpy models \citep{Levenson07}. 
The latter result was confirmed in the recent analysis of ground-based MIR spectra of nearby Seyferts 
presented in \citet{Gonzalez13}. The host galaxies of the Seyfert nuclei with the largest nuclear 9.7 \micron~apparent optical depths 
show nuclear dust lanes at optical wavelengths, are highly inclined and/or are part of a merger. 
The dust associated to these galaxy properties is likely contributing to the deep silicate features observed. 
A similar conclusion was reached by \citet{Goulding12} for a sample 
of Compton-thick Seyfert galaxies using data from the Spitzer Infrared Spectrograph (IRS) and probing kpc scales.

The previous works are useful for testing the models and constraining the torus parameters of AGN, 
considering that with current instrumentation we cannot obtain direct observations of the torus itself. The problem is that, 
when fitting SEDs with torus models, we normally use the IR data available in the literature for our targets. This implies that, 
for different objects of a given sample, the data were obtained with distinct filters and instruments, 
and the sampling of the SEDs may not be the same. Distinct SEDs of the same target can provide different fitting results, 
and for that reason, it is important to analise how IR observations at various wavelengths constrain the torus 
model parameters. This analysis is important because, to isolate as much as possible the torus emission, subarcsecond resolution 
data from ground-based 8-10 m class telescopes or the 
{\it Hubble Space Telescope (HST)} are necessary, and their over-subscription factors are very high. 

{\it BayesClumpy} \citep{Asensio09} is a computer program that can be used for the fast synthesis of 
SEDs emerging from the clumpy dusty torus models of Nenkova. These fast synthesis 
capabilities are used in a Bayesian scheme for carrying out inference over the model parameters for observed SEDs.
In the latest version of {\it BayesClumpy}\footnote{https://github.com/aasensio/bayesclumpy}, the inference can 
be done either using neural network interpolation (as described in \citealt{Asensio09}) or multilinear 
interpolation in the full database. After running different tests, we recommend the latter interpolation 
to the users for fitting the clumpy torus models of Nenkova. 
In addition, we have recently incorporated Bayesian adaptive exploration (BAE) in {\it BayesClumpy}
to predict which photometric filter is needed 
to maximize the expected utility \citep{Asensio13}. In other words, we used the information already present in the data
and clumpy torus models to evaluate the optimum next observation 
that maximizes the constraining power of the new observed photometric point.

Here we use BAE to study the SED sensitivity (including photometric and spectroscopic IR data)
to the six clumpy torus parameters of Nenkova. We also propose the optimum filterset required 
to constrain these parameters, using average and individual Sy1 and Sy2 SEDs of a small sample of nearly face-on, undisturbed 
Seyfert galaxies without prominent nuclear dust lanes. Although the Nenkova models are generic and widely used by the 
community, we note that some of the parameters that describe them might vary from model to model 
\citep{Honig06,Schartmann08,Stalevski12}. Besides, distinct models often treat differently the radiative transfer problem.
In fact, a given SED can be reproduced with a different set of two distinct model parameters. 
Thus, the results from SED fitting with torus models should be interpreted in the context of the chosen model only, as the 
resulting parameters are just a possible description of the geometry and properties of the torus, 
provided that the data are dominated 
by torus emission. See \citet{Honig13} for further discussion on SED fitting caveats. 
In summary, the results presented here are only valid, in principle, for the clumpy torus models of Nenkova. 
Throughout this paper we assume a cosmology with H$_0$ = 70 km s$^{-1}$ Mpc$^{-1}$, $\Omega_m$ = 0.27, 
and $\Omega_{\Lambda}$ =0.73.

\section{Sample selection}
\label{sample}

This work aims to study the sensitivity of different IR wavelengths to the clumpy torus parameters
of Nenkova. We then need a sample of nearby Seyfert galaxies whose SEDs and spectra can be accurately reproduced with 
these models. Thus, we selected three Sy1 and three Sy2 galaxies from \citealt{Alonso11} and 
\citet{Gonzalez13}
hosted in nearly face-on galaxies (inclination $\la45\degr$) without prominent nuclear dust lanes and which are not part of merging 
systems\footnote{The only exception 
is the Sy1.5 NGC\,3227, which is interacting with its companion galaxy NGC\,3226 \citep{Mundell04}. However, there are no
signs of morphological disturbance in its nuclear region, and the apparent depth of the 9.7 \micron~silicate feature is low 
($\tau_{9.7}<0.4$; \citealt{Gonzalez13}).}. 
The galaxies are NGC\,1068, NGC\,3081 and NGC\,5643 (Sy2s) and NGC\,1365, NGC\,3227 and NGC\,4151 (Sy1s) and they are at distances 
smaller than 19 Mpc, with the exception of NGC\,3081 (see Table \ref{sources}). This implies that, at the average resolution of 
ground-based N-band observations ($\sim$0.3-0.4\arcsec), we are probing the central 20-30 parsecs ($\sim$55 pc for NGC\,3081).

\begin{table}
\centering
\footnotesize
\begin{tabular}{lcccc}
\hline
\hline
Galaxy & Seyfert & Distance & Scale & Inclination \\
 & type & (Mpc) & (pc~arcsec$^{-1}$) & (deg) \\
\hline
NGC\,1365    &  1.8   &   18.6 (a) &  90  &  46 (g) \\  
NGC\,3227    &  1.5   &   17.0 (b) &  82  &  46 (h) \\ 
NGC\,4151    &  1.5   &   13.0 (c) &  63  &  31 (h) \\
NGC\,1068    &  2     &   14.4 (d) &  70  &  28 (h) \\  
NGC\,3081    &  2     &   32.5 (e) &  158 &  31 (i) \\ 
NGC\,5643    &  2     &   16.9 (f) &  82  &  29 (j) \\
\hline
\end{tabular}
\caption{Basic galaxy data. Inclination angle = 0$\degr$ corresponds to face-on galaxies.
References: (a) \citet{Madore98}; (b) \citet{Garcia93}; (c) \citet{Radomski03}; (d) \citet{Bland-Hawthorn97}; (e) \citet{Tully88};
(f) \citet{Gonzalez13}; (g) \citet{Jorsater95}; (h) \citet{Hunt92}; (i) \citet{Buta98}; (j) \citet{Jungwiert97}.}
\label{sources}
\end{table}

\section{Observations}

\subsection{Photometry}
\label{phot}

The NIR and MIR nuclear fluxes of the six galaxies considered here are reported in Tables \ref{nir} and \ref{mir}. 
For the galaxies NGC\,1068, NGC\,3081, NGC\,3227 and NGC\,4151, the data are the same reported
in \citealt{Alonso11} and \citealt{Ramos11}, and for NGC\,1365, in \citet{Alonso12}, with the following exceptions: 

In the case of NGC\,3227, we discarded the 8.99 \micron~VISIR/VLT flux from \citet{Honig10} and the 11.2 \micron~MICHELLE/Gemini 
North flux from \citealt{Ramos09} because they appear contaminated by [Ar III] and PAH 11.3 \micron~emission, respectively.  
In the NIR, we considered the L and M-band fluxes of NGC\,3227 and NGC\,4151 \citep{Ward87} as upper limits, because
of the limited angular resolution of the images (from NSFCam/IRTF in the case of the L-band and from IRCam3/UKIRT for 
the M-band observations). For NGC\,1068, we used the nuclear K-band flux reported by \citet{Thatte97} and 
we considered the H-band flux from \citet{Alonso01} as an upper limit, as it does not match the SED shape
and it possibly includes some level of contamination from extended emission \citep{Prieto10}. 
We refer the reader to \citealt{Alonso11} and \citealt{Ramos11} for a detailed description of the NIR and MIR 
observations. 

\begin{table*}
\centering
\begin{tabular}{lcccccc}
\hline
\hline
Galaxy & \multicolumn{5}{c}{NIR flux densities (mJy)} & Ref(s).  \\
& J band & H band & K band & L band & M band &  \\
\hline
NGC\,1365  & \dots         & 8.3$\pm$0.8  & \dots        & \dots       &  \dots        & 1      \\
NGC\,3227  & \dots         & 7.8$\pm$0.8  & 16.6$\pm$1.7 & $\leq$47    &  $\leq$72     & 2,3    \\   
NGC\,4151  & 69$\pm$14     & 104$\pm$10   & 177$\pm$18	 & $\leq$325   &  $\leq$449    & 4,3    \\
NGC\,1068  & 9.8$\pm$2.0   & $\leq$98     & 190$\pm$28   & 920$\pm$138 &  2270$\pm$341 & 5,6,7   \\  
NGC\,3081  & \dots         & 0.22$\pm$0.13& \dots  	 & \dots       &  \dots        & 8      \\   
NGC\,5643$^a$ &  \dots     & \dots        & $\leq$12.2   & \dots       &  \dots        & 9      \\ 
\hline
\end{tabular}
\caption{High spatial resolution NIR fluxes. (a) We included in the SED the 6 \micron~flux, {\bf of 61 mJy,} from ISOCAM/ISO reported by \citet{Lutz04} as an upper limit.
All the J, H and K fluxes are from NICMOS/HST observations (F110W, F160W and F222M filters) except the K-band fluxes of NGC\,1068 and NGC\,5643, which are
from SHARP/NTT and SOFI/NTT respectively. L and M fluxes are from COMIC/3.6 m ESO telescope in the case of NGC\,1068, 
and from NSFCam/IRTF (L) and IRCam3/UKIRT (M) for NGC\,3227 and NGC\,4151.
References: (1) \citet{Carollo02}; (2) \citet{Kishimoto07}; (3) \citet{Ward87}; (4) \citet{Alonso03}; (5) \citet{Alonso01}; (6) \citet{Thatte97};
(7) \citet{Marco00}; (8) \citet{Quillen01}; (9) this work.}
\label{nir}
\end{table*}						 

\begin{table*}
\centering
\footnotesize
\begin{tabular}{lcccccccc}
\hline
\hline
Galaxy & \multicolumn{4}{c}{MIR flux densities (mJy)} & Ref. & \multicolumn{2}{c}{N-band spectra} & Ref. \\
& N band & Filter & Q band & Filter & & Instrument & Slit (\arcsec) & \\
\hline
NGC\,1365     & 203$\pm$30		       & T-ReCS/Si-2		   &  818$\pm$204    & T-ReCS/Qa	     &  1  & T-ReCS   &  0.35 & 7      \\ 
NGC\,3227     & 320$\pm$48		       & VISIR/PAH2$_-$2	   & \dots	     & \dots		     &  2  & VISIR    &  0.75 & 2      \\ 
NGC\,4151     & 1320$\pm$200		       & OSCIR/N		   & 3200$\pm$800    & OSCIR/IHW18	     &  3  & MICHELLE &  0.36 & 8      \\  
NGC\,1068     & 10000$\pm$1500  	       & MIRTOS/8.72 \micron	   & 21800$\pm$5450  & MIRTOS/18.5 \micron   &  4  & MICHELLE &  0.36 & 9      \\
NGC\,3081     & 83$\pm$12		       & T-ReCS/Si-2  	           & 231$\pm$58      & T-ReCS/Qa	     &  3  & T-ReCS   &  0.65 & 7      \\   
     	      & 161$\pm$24	               & VISIR/PAH2		   & \dots	     & \dots		     &  5  & \dots    &  \dots& \dots  \\   
NGC\,5643     & 287$\pm$43		       & VISIR/PAH2$_-$2	   & 790$\pm$200     & T-ReCS/Qa	     & 2,6 & T-ReCS   &  0.35 & 7      \\ 
\hline
\end{tabular}
\caption{High spatial resolution MIR fluxes from T-ReCS/Gemini South, OSCIR/Gemini North, VISIR/VLT and MIRTOS/Subaru. 
References: (1) \citet{Alonso12}; (2) \citet{Honig10}; (3) \citealt{Ramos09}; (4) \citet{Tomono01}; (5) \citet{Gandhi09}; (6) this work; (7) \citet{Gonzalez13}; 
(8) \citealt{Alonso11}; (9) \citet{Mason06}. 
}
\label{mir}
\end{table*}

The Seyfert 2 galaxy NGC\,5643 was observed in May 2012 with the MIR camera/spectrograph T-ReCS 
\citep{Telesco98} on the Gemini-South telescope. The narrow Qa filter was employed 
($\lambda_c$=18.3 \micron, $\Delta\lambda$=1.5 \micron) and the 
angular resolution was 0.5\arcsec~at 18.3 \micron, as measured from the observed Point Spread
Function (PSF) star. The data were reduced using the {\it RedCan} pipeline \citep{Gonzalez13}, and the nuclear flux 
obtained from scaling the flux of the PSF star to the peak of galaxy emission, as described in \citealt{Ramos11}. The nuclear
Q-band flux of NGC\,5643, together with a 11.88 \micron~flux from \citet{Honig10} obtained with VISIR/VLT \citep{Lagage04}, 
are reported in Table \ref{mir}. The galaxy was also observed with T-ReCS using the narrow Si-2 filter 
($\lambda_c$=8.74 \micron, $\Delta\lambda$=0.78 \micron), but we did not include the Si-2 nuclear flux in the SED 
because it does not match the SED 
shape of the $>$10 \micron~points and the spectrum, as the flux calibration in this filter is uncertain\footnote{The galaxy 
was observed following the sequence: 1) PSF star at 8.7 \micron; 2) PSF star at 18.3 \micron; 3) NGC\,5643 at 8.7 \micron; 
4) NGC\,5643 at 18.3 \micron. PSF star observations 
in each filter should be made immediately prior to or after each galaxy observation to accurately sample 
the image quality and provide reliable flux calibration.}.

In the NIR, NGC\,5643 was observed with SOFI/NTT in June 1999 in the Ks filter with a seeing of FWHM = 1.6\arcsec. 
See \citet{Grosbol04} for further details on the SOFI observations. We performed aperture photometry on the Ks-band 
image of the galaxy in an aperture of 1 arcsec radius, and considered the resulting flux as an upper limit because 
of the image resolution (see Table \ref{nir}). Finally, we also included the ISOCAM/ISO 6 \micron~flux reported in 
\citet{Lutz04} in the SED as an upper limit.  

\subsection{Spectroscopy}
\label{spec}

In addition to the NIR and MIR photometry reported in Table \ref{nir}, we compiled N-band spectra (8--13 \micron) 
for all the galaxies considered here. Those spectra were obtained with the VISIR, T-ReCS and MICHELLE instruments, 
using the slit widths indicated in Table \ref{mir}. The T-ReCS spectra are from \citet{Gonzalez13}, and were reduced 
using the {\it RedCan} pipeline (described in detail in the latter paper). The MICHELLE spectra are from \citet{Mason06} and 
\citealt{Alonso11} and the VISIR spectrum of NGC\,3227 is from \citet{Honig10}. 

We note that the apparent strength of the nuclear 9.7 \micron~optical depth measured for the six galaxies considered here is low
($\tau_{9.7}<1$; \citealt{Honig10b,Alonso11,Gonzalez13}) and thus, it can be reproduced with clumpy torus models, 
as explained in the Introduction.

\subsection{Average Sy1 and Sy2 photometry and spectroscopy}
\label{average}

Since we are looking for general properties of Sy1 and Sy2 SEDs, 
we constructed average Sy1 and Sy2 templates using our small sample of nearby, nearly face-on and undisturbed Seyfert galaxies. 

The best-sampled SEDs, those of NGC\,1068 and NGC\,4151, define the wavelength grid, from 
1.6 to 18.2 \micron~(see Table \ref{average_seds}).
This grid includes seven photometric data points, and we interpolated nearby measurements of the other galaxies onto this scale, 
excluding upper limits\footnote{For NGC 5643 we could not obtain interpolated NIR fluxes, as we only have an upper
limit of the K-band flux.}. 
The average Sy1 and Sy2 SEDs, normalised at 11.88 \micron, are reported in Table \ref{average_seds}.
Errors correspond to the nominal percentages considered for NIR and MIR nuclear flux measurements (15\% for all of them but
Q-band fluxes, for which we consider 25\% errors). See \citealt{Ramos11} for further details on IR flux uncertainties. 

\begin{figure}
\centering
\includegraphics[width=9cm]{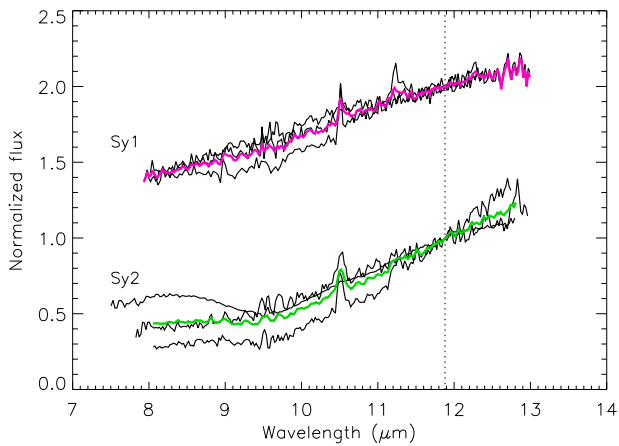}
\caption{Individual (black lines) and stacked Sy1 (top; magenta thick line) and Sy2 (bottom; green thick line) 
normalised rest-frame spectra. The Sy1 spectra have been shifted in 
the Y-axis for clarity. The vertical dotted line indicates the wavelength chosen for the normalisation. 
\label{stack}}
\end{figure}

We also stacked the individual Sy1 and Sy2 spectra, once normalised at 11.88 \micron, using the IRAF\footnote{IRAF is distributed by 
the National Optical Astronomy Observatory, which is operated by the Association of Universities for Research in Astronomy (AURA) 
under cooperative agreement with the National Science Foundation.}
task {\it scombine} with the ``average'' option to combine the data. The individual and stacked spectra are shown in 
Figure \ref{stack}.

\begin{figure}
\includegraphics[width=9cm]{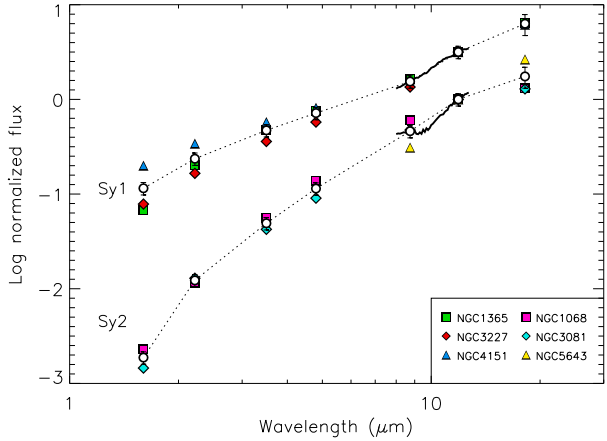}
\caption{Individual (filled symbols) and average (open circles) Sy1 and Sy2 interpolated rest-frame SEDs. Sy1 SEDs have been shifted in 
the Y-axis for clarity. The average 8--13 \micron~spectra have been resampled to $\sim$50 points, as in
\citet{Alonso13}. 
\label{sy1sy2}}
\end{figure}

Finally, in Figure \ref{sy1sy2} we show the average Sy1 and Sy2 photometry and spectroscopy, as well as the
interpolated photometric data for the individual galaxies. All the data were normalised at 11.88 \micron, 
and the Sy1 data shifted in the Y-axis for clarity. The main differences between the two Seyfert types are the flatter 
NIR slope of the Sy1s as compared to the Sy2s, and 
the shallow 9.7 \micron~silicate absorption in the Sy2 spectrum, which is absent in the Sy1 data. 

In \citealt{Ramos11} we reported average Sy1 and Sy2 SEDs as well, but constructed from a larger sample and 
considering photometry only. 
By comparing the latter with Table \ref{average_seds}, we find no difference between the Sy1 average 
SEDs, and only a slightly higher H-band flux in the case of the average Sy2 SED considered here.
However, this flatter NIR slope is compatible with the individual Sy2 SEDs in \citealt{Ramos11}.

\begin{table}
\centering
\large
\begin{tabular}{ccc}
\hline
\hline
$\lambda_c$ & Sy1 SED & Sy2 SED \\
(\micron) & (mJy) &  (mJy)\\
\hline
1.60    &  0.036$\pm$0.005    & 0.0019$\pm$0.0003 \\
2.22    &  0.075$\pm$0.011    & 0.012$\pm$0.002 \\
3.50    &  0.15$\pm$0.02      & 0.05$\pm$0.01 \\
4.80    &  0.23$\pm$0.03      & 0.11$\pm$0.02 \\
8.74    &  0.49$\pm$0.07      & 0.46$\pm$0.07 \\
11.88   &  1.00$\pm$0.15      & 1.00$\pm$0.15 \\ 
18.17   &  1.99$\pm$0.50      & 1.75$\pm$0.44 \\ 
\hline      
\end{tabular}						 
\caption{Normalised average Sy1 and Sy2 SEDs constructed by using the individual photometric and spectroscopic data of NGC\,1365, NGC\,3227, 
NGC\,4151 (Sy1), NGC\,1068, NGC\,3081 and NGC\,5643 (Sy2).}
\label{average_seds}
\end{table}

\section{Results}

\subsection{Clumpy torus models and Bayesian approach}
\label{bayesian}

\begin{figure}
\includegraphics[width=8cm,angle=90]{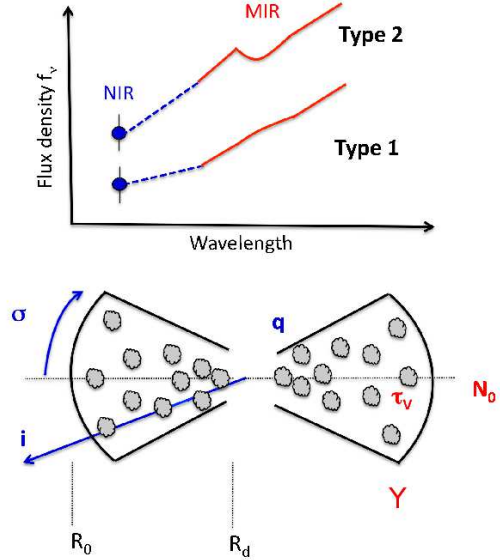}
\caption{Top: schematic view of a typical Sy1 IR SED (flat IR slope and absent silicate feature) and same for the 
Sy2 (steeper IR slope and silicate feature in shallow absorption). Bottom: schematic representation of a clumpy torus. 
Parameters that depend on the MIR--to--NIR ratio are represented in blue ($\sigma$, $i$ and $q$) and those that require MIR 
spectroscopy in red ($\tau_V$, $Y$ and N$_0$).
\label{squeme}}
\end{figure}

The clumpy models of Nenkova hold that the dust
surrounding the central engine of an AGN is distributed in clumps, with a radial extent $Y = R_{o}/R_{d}$. 
$R_{o}$ and $R_{d}$ are the outer and inner radius of the toroidal distribution, respectively (see bottom part of Figure \ref{squeme}). 
The inner radius is defined by the dust sublimation temperature ($T_{sub} \approx 1500$ K; \citealt{Barvainis87}),
with $R_{d} = 0.4~(1500~K~T_{sub}^{-1})^{2.6}(L / 10^{45}\,\mathrm{erg ~s^{-1}})^{0.5}$ pc.  
Within this geometry, each clump has the same optical depth ($\tau_{V}$), measured at 5500 \AA.
The average number of clouds along a radial equatorial ray is $N_0$. The radial density profile is a
power-law ($\propto r^{-q}$) and $\sigma$ characterizes the angular distribution of the clouds, which has
a smooth edge. Finally, the number of clouds along the line of sight (LOS) 
at an inclination angle $i$ is $N_{LOS}(i) = N_0~e^{(-(i-90)^2/\sigma^2)}$. 
We refer the reader to \citet{Nenkova08a,Nenkova08b} for a more detailed description of the clumpy model parameters. 

In the Bayesian scheme, we can specify a-priori information about the model
parameters. We consider the priors to be truncated uniform distributions for the six model parameters 
in the intervals reported in Table \ref{parametros}. Therefore, we give the same weight
to all the values in each interval. The results of the fitting process of the IR SEDs (see Tables \ref{nir} and 
\ref{mir}) with {\it BayesClumpy} are the posterior distributions for the six free parameters that describe the models. 
When the observed data introduce sufficient information into the fit, 
the resulting posteriors will clearly differ from the input uniform distributions, 
either showing trends or being centered at certain values within the intervals considered. A
detailed description of the Bayesian inference applied to clumpy models can be found in \citet{Asensio09}. 
In addition, to see different examples of the use of clumpy model fitting to IR SEDs using {\it BayesClumpy}, see
\citealt{Ramos09}, \citealt{Ramos11} and \citealt{Alonso11}.

\begin{table}
\centering
\begin{tabular}{ll}
\hline
\hline
Parameter &  Interval \\
\hline
Width of the angular distribution of clouds ($\sigma$) & [15\degr, 75\degr]	       \\
Radial extent of the torus ($Y$)                       & [5, 100]		       \\
Number of clouds along equatorial direction ($N_0$)    & [1, 15]		       \\
Power-law index of the radial density profile ($q$)    & [0, 3] 		       \\extinc
Inclination angle of the torus ($i$)               & [0\degr, 90\degr]	       \\
Optical depth per single cloud ($\tau_{V}$)            & [5, 150]		       \\ 
\hline      
\end{tabular}						 
\caption{Clumpy model parameters and intervals considered as uniform priors. For the Sy1s, we also considered foreground extinction as 
an additional parameter (A$_V\leq$ 3 mag, based on the values of the invidual galaxies; \citealt{Alloin81,Ward87b,Mundell95}). 
}
\label{parametros}
\end{table}

We fit the Sy2 SED considering reprocessed torus emission and no foreground extinction, since NGC\,1068, 
NGC\,3081 and NGC\,5643 are undisturbed, nearly face-on galaxies without prominent nuclear dust lanes. On the other hand, 
for the Sy1 SED, which is the average of Seyfert 1.5 and 1.8 SEDs, we need to consider the AGN 
direct emission (see \citealt{Ramos09} and \citealt{Ramos11}). 
Following \citet{Nenkova08a}, we include a piecewise power-law distribution \citep{Rowan95}
in addition to the torus reprocessed emission. 

In the case of the Type-1 Seyferts, we also allowed for a small amount
of foreground extinction, separated from the torus, to redden the direct AGN emission. We consider this
foreground extinction as an additional parameter in the fit (A$_V^{for}\leq$ 3 mag, after considering the individual 
A$_V$ values reported in the literature for the individual Sy1s considered here) and we
use the IR extinction law of \citet{Chiar06} in the range $\sim$1--35 \micron, which accounts for 
the two silicate features at 9.7 and 18 \micron. Foreground extinctions smaller than $\sim$5 mag have negligible effect
in the torus parameters derived from the Sy2 fits \citep{Alonso03}, but they can be significant for Sy1s, whose 
NIR SEDs have an important contribution from AGN direct emission~\citep{Ramos09}. 


Finally, we resampled all the 8--13 \micron~spectra to $\sim$50 points, following the same methodology as in \citet{Alonso13}.

\subsection{Bayesian adaptive exploration}

\begin{figure*}
\centering
\includegraphics[width=17cm]{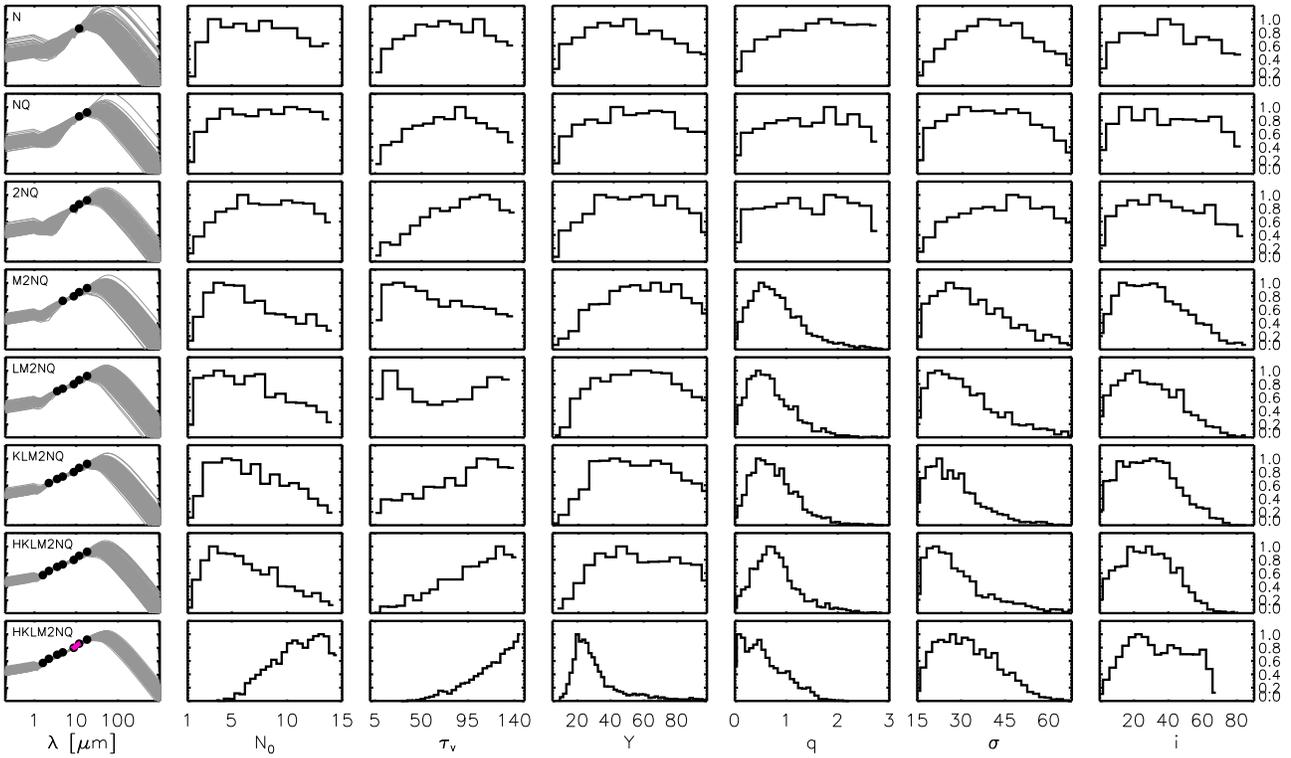}
\caption{BAE experiment for the average Sy1 SED. A new observed point from Table \ref{average_seds} is added in each step
(indicated with labels in the first column). This SED has N-band fluxes at 8.74 and 11.88 \micron, which are labelled as 
2N when they are both included in the fit.
The first column also shows the model SEDs obtained from the posterior distributions in each case. 
The next six columns represent the marginal posteriors of the clumpy parameters. The last row includes the
stacked Sy1 spectrum shown in Figure \ref{stack}. 
\label{sy1}}
\end{figure*}

\begin{figure*}
\centering
\includegraphics[width=17cm]{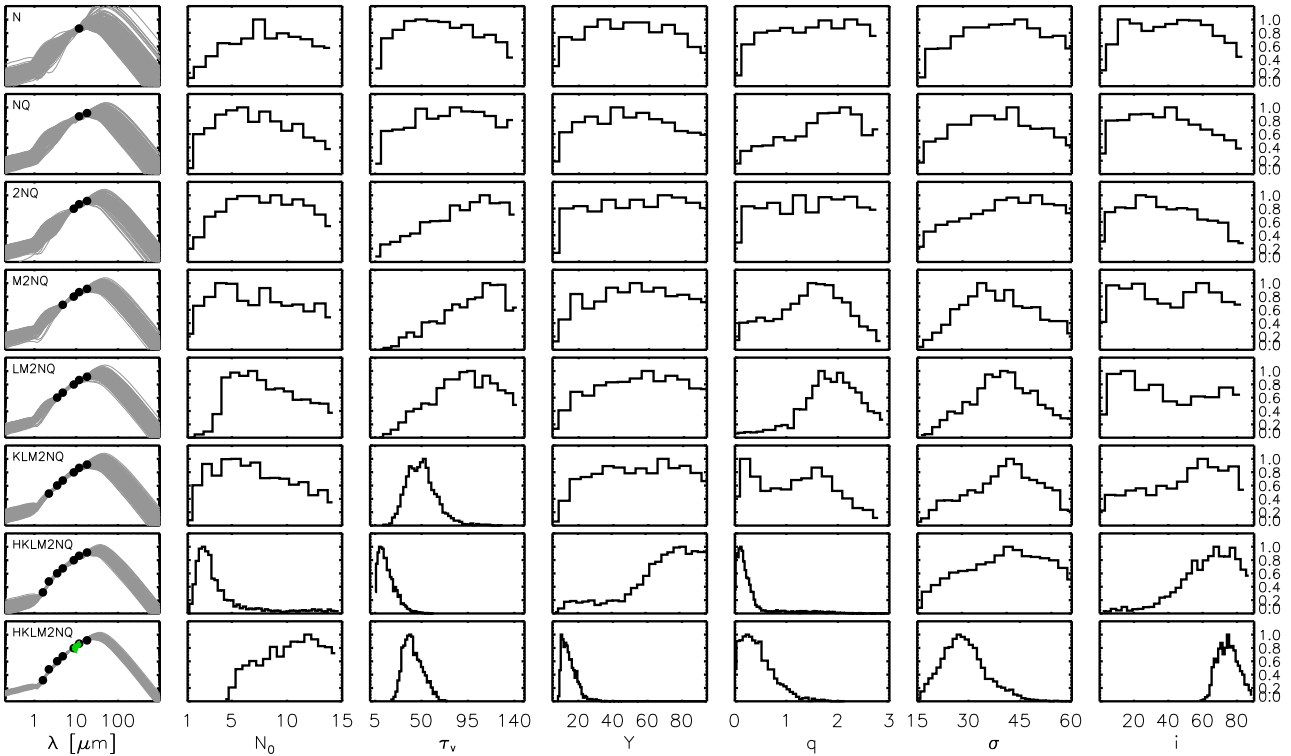}
\caption{Same as in Figure \ref{sy1}, but for the average Sy2 SED.  
\label{sy2}}
\end{figure*}

The aim of this work is to analise the SED sensitivity to the clumpy model parameters of Nenkova, 
rather than providing the best fitting results to IR SEDs, as we did in previous studies. 
To this end, we used BAE to evaluate the optimum 
next observation that maximizes the constraining power of the SED \citep{Asensio13}.
In Figures \ref{sy1} and \ref{sy2} we show the simulated iterated BAEs for the average Sy1 and Sy2 SEDs (photometry + 
spectroscopy), respectively, as we did in \citet{Asensio13} for NGC\,3081.

As a first step, we assumed that we only have one photometric point in the N-band (at 11.88 \micron; see Table \ref{average_seds}) 
and we let the BAE scheme select the following filters. {\it BayesClumpy} then uses that point to sample the posterior 
distribution and obtains marginal posteriors for the clumpy model parameters. The 11.88 \micron~point, together 
with the model SEDs sampled from the posterior, are shown in the first column of Figures \ref{sy1} and \ref{sy2}. 
The remaining panels display the marginal posteriors for the six clumpy model parameters.

In order to simulate the next experiment, we selected the next observation in the observed 
SED that has the largest expected utility, which in the two cases is the Q-band data point (18.2 \micron). 
The full process is then repeated until we have all the
observed points of the SED. In both Sy1 and Sy2 BAE experiments, the next observation was always
the closest in wavelength to the previous one: Q, Si-2, M, L, K and H. This is logical if we consider that the bulk 
of the torus emission peaks in the MIR (see \citealt{Asensio13}). 
Finally, we added the stacked and resampled N-band spectra. 
All the experiments performed for both the average Sy1 and Sy2 SEDs and the corresponding posteriors are shown in 
Figures \ref{sy1} and \ref{sy2}. We did the same experiment for the individual SEDs in order to confirm the results, 
but we do not include the figures here for the sake of simplicity. 



\subsubsection{Sy1}
\label{sy1section}

As described in Section \ref{bayesian}, we fitted the Sy1 SED with reprocessed torus emission, AGN direct emission, 
and allowed for a small amount of foreground extinction (A$_V\leq$ 3 mag). We show the results of the BAE experiment
in Figure \ref{sy1}.

When we only consider MIR photometry (i.e. 8 to 18 \micron~data; three top rows in Figure
\ref{sy1}), we cannot constrain any of the torus parameters (i.e. the posteriors do not 
significantly differ from the uniform priors). It is when we add M and L-band photometry 
that we start to see trends: small torus widths and inclination angles ($\sigma$ and $i$ respectively),
flat cloud distributions and low number of clumps ($q$ and $N_0$) become more probable (4th and 5th rows). 
We note that at this point, the silicate feature at 9.7 \micron~is predicted in emission in the fitted models.

When we include H and K-band data in the fit we observe the same trends in the posteriors, with the 
only exception of $\tau_V$, for which large values within the considered range are preferred (6th and 7th rows).
Finally, when we include the N-band spectrum in the fit, which shows a flat silicate feature, 
the number of clumps increases to reproduce it, and $Y$ appears well-constrained. Thus, from this experiment 
and the individual ones we can infer the following:

\begin{itemize}

\item Flat silicate features can be reproduced with relatively large number of clouds (N$_0\sim$10--15) 
and large optical depth ($\tau_V\sim$100--150). On the other hand, strong silicate emission is associated to 
optically thin dust (low N$_0$ and $\tau_V$; \citealt{Sturm05,Mason12}). Other combinations of clumpy torus parameters can 
produce these features (see Figure 16 in \citealt{Nenkova08b}), but here we discuss the results of experiments with real
observations.

\citet{Honig10b} claimed that silicate emission in type-1 AGN 
can be also associated with steeper cloud distributions (i.e. large values of $q$), in which the majority of the clumps 
are close to the active nucleus, and consequently hotter than in flat radial distributions (low values of $q$). The larger 
amount of hot dust produces a silicate feature in emission. From the fits of the average and individual Sy1 SEDs studied 
here (see Appendix \ref{appendixA}), which show absent silicate features, we obtain flat--to--intermediate 
radial profiles ($q$=0--1.5; i.e. cooler dust).

\item Adding NIR data (1 to 5 \micron) to the MIR SED constrains the torus width ($\sigma$) and the inclination angle of the torus ($i$). 
In the case of the Sy1s, which generally show flatter IR SEDs than Sy2s (see Figure \ref{sy1sy2}), it appears enough to add L or 
M-band photometry to the MIR data to reliably constrain $\sigma$ and $i$.

\item Another parameter that it is constrained after including L and/or M-band data in the fit is the index 
of the radial density profile ($q$), which defines the distribution of the clumps. \citet{Honig10b}
reported on a tight correlation between the MIR slope of Type-1 AGN (measured from linear fitting between the 7--8.5 \micron~and 
13.4--14.6 \micron~regions) and $q$, with flatter slopes
corresponding to redder MIR colors.

\item The $q$ parameter is closely related to the torus size ($Y$). In clumpy models, for steep radial density distributions 
($q$=2, 3), the SED is never sensitive to the outer torus extent. This is because the majority of the clouds are located
in the inner part of the torus, closer 
to the nucleus for all values of Y. In models with flatter radial profiles ($q$=0, 1), more clumps are located
farther from the central engine. These SEDs are more sensitive to $Y$, but only at wavelengths $\ga$20 \micron, according
to \citet{Nenkova08b}.

In the case of the Sy1 SED, we have a flat radial profile that allows us to constrain the torus size ($Y$) when 
we add the N-band spectrum. This means that it is possible to constrain the torus size by including a 
8--13 \micron~spectrum when we have a flat radial profile.
It is worth noting, however, that this torus size corresponds to the outer radius of the MIR emitting dust. 
Cooler dust emits in the far-IR (FIR) and indeed, data from the {\it Herschel Space Observatory} 
at 70--500 \micron~have proved to be useful for constraining $Y$ and $q$ (\citealt{Ramos11b,Alonso12}).

Finally, we find that including a small amount of foreground extinction in the fits (A$_V^{for}\leq$3 mag) 
does not have a significant effect in the torus parameters. However, for galaxies with 
A$_V^{for}\ga$5 mag, the effect of extinction has to be taken into account, not only for
Sy1s, but also for Sy2s, especially when trying to reproduce the 9.7 \micron~silicate feature and the NIR emission (see also
\citealt{Alonso11}). 
For this reason, we emphasize that it is important to either select AGN as those discussed here (free of nuclear dust lanes, 
undisturbed and face-on), or have accurate measurements of the nuclear foreground extinction,  
when attempting to derive torus parameters from SED fitting.

\end{itemize}

\subsubsection{Sy2}
\label{sy2section}

We performed the BAE experiment for the Sy2 SED considering torus reprocessed emission only, without foreground
extinction. 
If we look at Figure \ref{sy2}, we see that, as it happens with the Sy1 fit, none of the posteriors 
is really constrained when we only fit MIR photometry (three top rows). It is after introducing L and M-band 
photometry in the fit (4th and 5th rows) that the parameters start to show trends. However, when we include H and K-band data 
(6th and 7th rows) the majority of these trends change, differently to what happens to the Sy1 SED (see Figure \ref{sy1}). 
From this experiment and the individual ones we conclude the following:


\begin{itemize}

\item 
The models reproduce the silicate feature in absorption (see Figure \ref{sy1sy2}) with large number of clumps
(N$_0\sim$8--15) and intermediate optical depth of the clouds within the considered interval
($\tau_V\sim$50), which usually happens when fitting Sy2 SEDs (\citealt{Alonso11}). 
  
\item As in the case of the Sy1 fit, the MIR--to--NIR slope is sensitive to $q$, $\sigma$ and $i$, with 
higher MIR--to--NIR ratios associated, in general, with intermediate--to--edge-on 
views\footnote{In the clumpy torus squeme that we are
considering, face-on views correspond to $i=0\degr$ and edge-on to $i=90\degr$.}. 
In the case of the Sy2, however, we need data at $\lambda<$3 \micron~to reliably constrain $q$, 
$\sigma$ and $i$. The steeper NIR slope of Sy2 SEDs, as compared to those of Sy1s, 
requires to be sampled using the J, H and/or K-bands to reliably constrain these parameters.

The case of $\sigma$ appears not as clear-cut as those of $q$ and $i$ if we look at Figure \ref{sy2} only.
The corresponding posterior in the 7th column (i.e. before including the spectrum) is not well constrained.
However, if we consider the BAE experiments performed for NGC\,3081 and NGC\,5643, we find that 
we can reliably constrain $\sigma$ by including NIR data in the fit, without spectroscopy. 
We note that in the case of the Sy2 fit, we obtain a low value of
$\sigma$, a behaviour that is not common in Sy2s (see \citealt{Ramos11} and \citealt{Alonso11}). 
This is due to the slightly flatter NIR slope of the Sy2 SED, 
as compared to the average Sy2 SED in \citealt{Ramos11}. In fact, the NIR SED shape resembles that of NGC\,1068, 
whose fit produces similar posteriors (see Appendix \ref{appendixA}). 

\item Regarding $q$ and $Y$, we see roughly the same behaviour as in Figure \ref{sy1}: if we have a flat radial 
profile, as it is the case ($q<1$), we can reliably constrain $Y$ by fitting the spectrum in addition to the 
photometry, which efficiently reduces the parameter space. On the other hand, in the case of NGC\,3081, we
have a steep radial profile ($q>2$), and we cannot constrain $Y$ by including the spectroscopy 
(see Appendix \ref{appendixA}). 

\end{itemize} 




\subsubsection{SED sensitivity to the torus parameters}

From the BAE experiments performed in the previous sections for the average and individual 
Sy1 and Sy2 SEDs, we can conclude the following:

First, having a well sampled NIR/MIR SED, with photometric data only, and probing scales $\la$50 pc, 
it is possible to reliably constrain the torus width ($\sigma$), its inclination ($i$) and the radial profile of 
the clumps ($q$). The resulting posteriors are practically the same if we do and do not include N-band spectroscopy. 


Second, in addition to the NIR and MIR photometry, we need subarcsecond 8--13 \micron~spectroscopy to put 
realistic constraints on the torus extent ($Y$). If the cloud distribution is flat 
($q$=0, 1), the N-band spectrum efficiently reduces the number of models compatible with the MIR data 
(see bottom rows of Figures \ref{sy1} and \ref{sy2}) and we can constrain the extent of the MIR-emitting dust ($Y$).
On the other hand, if we have a steep radial profile ($q$=2, 3) the NIR/MIR SED is not sensitive to the 
outer torus extent, as the majority of the clumps are close to the nucleus, and we cannot put reliable constraints on $Y$. 
In any case, if we pursue a reliable estimate of the torus size, FIR data is required for probing the coolest 
dust within the torus.

We find that the depth of the 9.7 \micron~silicate feature is sensitive 
to the number of clumps and their optical depth (N$_0$ and $\tau_V$). Strong emission features 
are normally associated to optically thin dust (low N$_0$ and $\tau_V$). On the other hand, and 
according to the experiments performed here, flat/absent 
silicate features in Sy1 translate in large values of N$_0$ and $\tau_V$, while the 
shallow absorption features normally seen in Sy2 galaxies are generally reproduced with high values 
of N$_0$ and intermediate $\tau_V$ (see \citealt{Alonso11} for more examples of Sy2 fits including spectroscopy). 

The silicate feature is also dependent, although to a lesser extent, on the radial cloud distribution. Absent
silicate features in Sy1 and stronger absorptions in Sy2 are normally associated to flatter cloud distributions
(i.e. more clumps located farter from the active nucleus and consequently cooler; \citealt{Honig10b}). 

In \citealt{Ramos11}, we fitted NIR and MIR photometry only with the clumpy torus models of Nenkova, and the 
silicate features were generally predicted in emission. In those cases we obtained lower values
of $N_0$ from the fits. On the other hand, \citealt{Alonso11} included N-band 
spectroscopy in the fits, which in the case of the four Sy1 studied, showed absent silicate features. 
As it happens here, weak or absent silicate features can be 
reproduced with large number of clumps (N$_0\sim$10--15) and optical depths ($\tau_V\sim$100-150). 
For the Sy2s, which generally show shallow silicate absorption features, we find large values of $N_0$ 
($\sim$8--15) and intermediate $\tau_V$ ($\sim$50), as in \citealt{Ramos11} and \citealt{Alonso11}.

Thus, in general, {\it we conclude that we can constrain torus width, inclination and distribution of the clouds
with NIR and MIR photometry only, but MIR spectroscopy is necessary to restrict the posterior distributions of 
the number of clouds and their optical depth. The torus extent can also be constrained with MIR spectroscopy, 
but only in those cases where the clumps show a flat radial distribution.} 
This is summarised in Figure \ref{squeme}, where we represent the parameters that 
depend on the MIR--to--NIR ratio in blue and those that require MIR spectroscopy in red.

\subsection{Minimisation of the filterset}
\label{filter}

Looking at the evolution of the posteriors in Figures \ref{sy1} and \ref{sy2}, we have learnt that N-band spectroscopy 
and NIR photometry are necessary to constrain all the torus parameters. Now we can investigate the minimum 
filterset needed to obtain the same posterior distributions as those in the last row of Figures \ref{sy1} and \ref{sy2}.
Thus, we repeated the experiment for the average Sy1 and Sy2 SEDs, but starting with the N-band spectrum and 
the 11.88 \micron~point only. In this experiment, instead of selecting the next observation with the largest utility, we
tried different combinations of filters. The results are plotted in Figures \ref{sy1red} and \ref{sy2red}.

\subsubsection{Sy1 minimum filterset}

\begin{figure*}
\centering
\includegraphics[width=17cm]{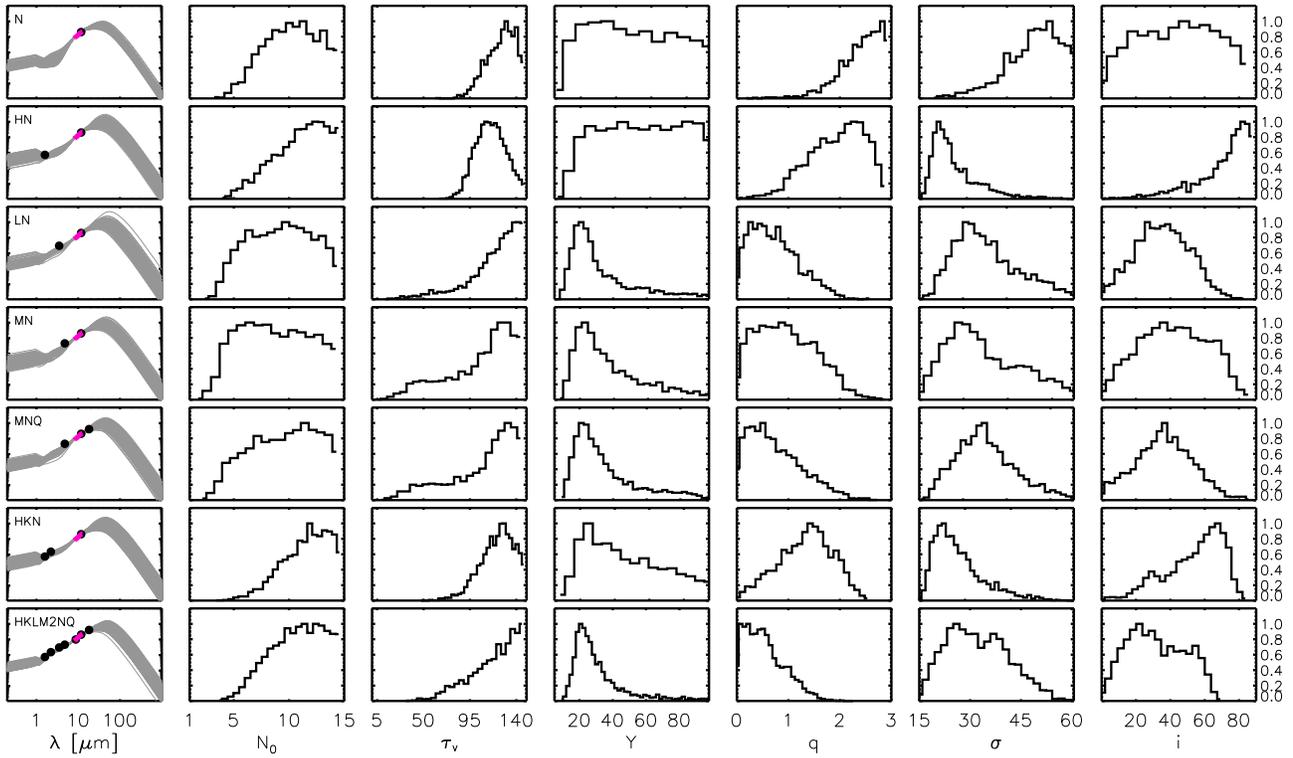}
\caption{Same as in Figure \ref{sy1}, but starting the Sy1 experiment with the N-band spectrum and 11.88 \micron~point, and 
choosing different combinations of filters. 
\label{sy1red}}
\end{figure*}

In the case of the Sy1 SED, the N$_0$ and $\tau_V$ posteriors in the first row of Figure \ref{sy1red}
are practically identical to those obtained when we fit the 
whole SED (bottom row). This is because the N-band spectrum is sensitive to N$_0$ and $\tau_V$, 
and in this case, the absent silicate feature itself rules out low values of these parameters 
(see Section \ref{sy1section}). In the second, third and fourth rows of Figure \ref{sy1red} we 
added one NIR data point to the fit (H, L and M respectively). Including H-band data (or K-band, as we also checked)
constrains the $\sigma$ and $i$ posteriors, as expected. If we include L- or M-band photometry instead, 
the resulting posteriors are exactly as those in the last row of Figure \ref{sy1red}. In the fifth 
row we checked that adding Q-band data to the M + N-band spectrum does not 
have any significant effect on the fit. Finally, in the sixth row we fitted the combination of H+K+N-band data, which also 
produces practically the same posteriors as in the bottom row, except for $q$ and $i$. 

The lack of constraining power of the Q-band data in the fit, when used in combination with N-band spectroscopy, 
is noteworthy. 
What is happening is that the 8--13 \micron~spectrum itself restricts the parameter space in the MIR, making it 
rather unnecessary to include 18 \micron~data. 
This effect is clearly illustrated in the top left panels of Figures \ref{sy1} and \ref{sy1red}. The number of models compatible
with the 11.88 \micron~photometry only in the first row of Figure \ref{sy1} are far more than those in the same row of Figure \ref{sy1red},
with the spectrum included. 
According to \citet{Nenkova08b}, we need data at wavelengths $\ga$20 \micron~to constrain torus outer extent ($Y$), 
but here we find that N-band spectroscopy reduces more efficiently the parameters space, and it is the 
addition of the spectrum, in combination with NIR photometry, what constrains the torus size when we have flat radial
profiles ($q$=0, 1; see the two last rows of Figures \ref{sy1} and \ref{sy2}).

As we are fitting average SEDs, and different SED shapes are sensitive to different torus properties, 
we have repeated the previous experiment for the individual galaxies. In particular, we have chosen NGC\,3227 
and NGC\,4151, whose SED shapes are different between them and well-sampled (see Figure \ref{sy1sy2}). The results of 
these experiments are shown in Figures \ref{ngc3227} and \ref{ngc4151} in Appendix \ref{appendixA}. The 
individual fits confirm the results obtained for the average Sy1 SED, but suggest a slightly different 
minimum filterset. In both cases, the combination of H+K+N-band data produces exactly the same posteriors 
as when fitting the whole dataset (see fourth and third rows of Figures \ref{ngc3227} and \ref{ngc4151} respectively). 
Unfortunately, we only have L and M-band upper limits for NGC\,3227 and NGC\,4151, so we cannot check if the 
combination of L/M and N-band data produces the same results. 

Thus, considering the average and individual Sy1 fits, the combinations of 8--13 \micron~spectrum 
and M- or H+K-band photometry appear to be the minimum necessary to constrain the torus parameters of Sy1 galaxies. 
If possible, we encourage the potential users of torus models to use a combination of 
two NIR data points (among J, H and K) in addition to M-band photometry + N-band spectroscopy 
for accurately constraining the torus parameters of Sy1 galaxies, although the minimum filterset
for individual SEDs appears to be a 8--13 \micron~spectrum + M-band photometry.

\subsubsection{Sy2 minimum filterset}

\begin{figure*}
\centering
\includegraphics[width=17cm]{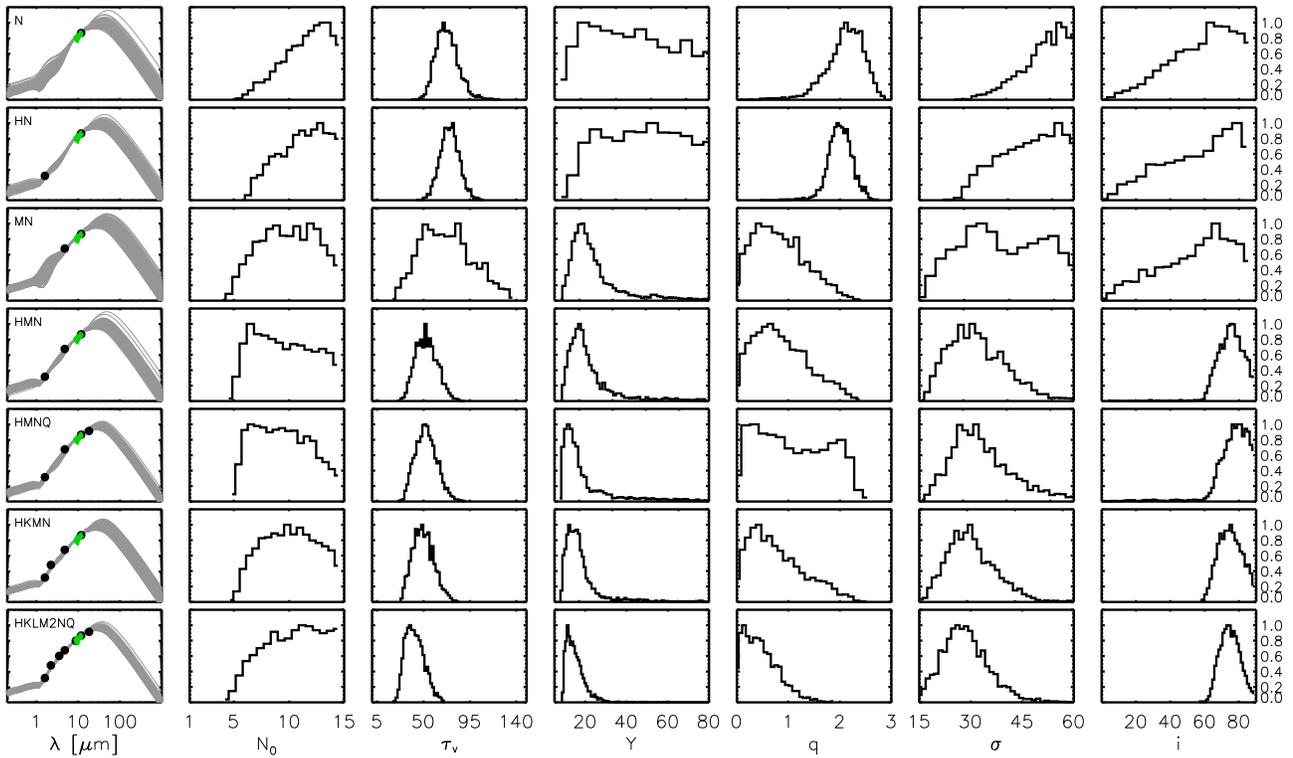}
\caption{Same as in Figure \ref{sy1red}, but for the Sy2 average SED and choosing different combinations of filters. 
\label{sy2red}}
\end{figure*}

The experiment with the average Sy2 SED produces similar results. In the second and third rows of Figure
\ref{sy2red} we added H and M-band data, respectively, to the N-band data. As for the Sy1 fit, the combination 
of M and N-band data (3rd row) produces posteriors which are similar to those in the last row of Figure \ref{sy2red}.
However, it is when we fit H+M+N-band data that we obtain the same results as when we fit the whole dataset 
(4th row). 
This is likely due to the steeper slope of Sy2 SEDs, which is better  
defined by the H+M+N-band data. We obtain the same results when using H+M+N+Q-band and H+K+M+N-band data (5th and 
6th rows). 

As we did for the Sy1 galaxies, in Appendix \ref{appendixA} we show the results from the individual 
SED fits of NGC\,1068 and NGC\,3081 (see Figures \ref{ngc1068} and \ref{ngc3081}). In the case of NGC\,1068, all the posteriors  
but $\sigma$ and $Y$ are well-constrained when we fit J+M+N-band data (4th row). In this case we have a steep radial 
profile ($q\sim$2) and thus, the SED is not sensitive to $Y$, as explained in Section \ref{sy1section}. 
In the fifth row of Figure \ref{ngc1068} we confirmed the lack of constraining power of the Q-band data
when used in combination with N-band spectroscopy.
Finally, the fit of J+K+M+N-band data (6th row) produces the same $\sigma$ posterior
as in the bottom row of Figure \ref{ngc1068}. The SED of NGC\,1068 is peculiar, showing a  
NIR bump that it is reproduced with a small torus width, more characteristic of Sy1 galaxies in 
general (\citealt{Ramos11}). In any case, it is noteworthy that the J-band data seems to work better in this case 
than the H-band data for constraining $\sigma$ and $i$. 

The fit of the whole NIR+MIR dataset of NGC\,3081, which is shown in the fourth row of Figure \ref{ngc3081}, 
produces posteriors much more typical of Sy2 galaxies (\citealt{Ramos11}). In this case, we just need to fit
H+N-band data (2nd row) to obtain the same posteriors as with the whole dataset. In fact, using N-band 
spectroscopy only we have practically the same
result (top row) with slightly broader posteriors. In the case of NGC\,3081, we also have nuclear FIR fluxes 
from the {\it Herschel Space Observatory} that we published in \citet{Ramos11b}. 
In the 5th, 6th and 7th rows of Figure \ref{ngc3081}, we added PACS 70, 100 and 160 \micron~data to the fit respectively. 
Including the FIR data points only affects the $Y$ posterior, which has a maximum of probability towards larger
values (bottom row). In this case we have a very steep radial profile ($q$=2-3), and thus, only with FIR data it 
is possible to have a weak constraint on the torus size, which is this case is relatively large, as we 
are probing cooler dust within the torus.   

Considering the average and individual Sy2 fits discussed here, the minimum combination of data needed to 
reliably constrain the torus parameters is H+M-band photometry + N-band spectroscopy, although in some cases, 
the use of J+K+M-band fluxes, in addition to the 8--13 \micron~spectrum, is necessary to correctly constrain the 
torus width. Thus, whenever possible, we encourage the reader to use two NIR data points (preferably including J-band data)
in addition to M-band photometry and N-band spectroscopy. This filterset ensures a reliable 
determination of all the torus parameters of Sy2 galaxies. 

The difference between the minimum filtersets of Sy1s and Sy2s is likely related to the AGN contribution to the SED. 
In the case of the Sy2s, whose IR SEDs can be reproduced with torus emission only, 
the parameters are very sensitive to different SED shapes. On the other hand, for the Sy1s, the intrinsic AGN emission is 
strong in the NIR, and thus the parameters are less sensitive to SED variations.

Summarising, based on our analysis of average and individual IR SEDs of Seyfert galaxies, we conclude that the minimum 
combination of IR data necessary 
to constrain the torus parameters is M-band photometry + 8--13 \micron~spectroscopy for Sy1 galaxies, 
and H+M-band photometry + 8--13 \micron~spectroscopy for Sy2 galaxies. However, {\it to reliably constrain all the torus parameters
of both Sy1 and Sy2 galaxies, independently of SED shape, we recommend the use of J+K+M-band photometry and 
N-band spectroscopy}\footnote{Any combination of two NIR filters (among J, H and K) should be valid when added to the M+N-band data, 
but J+K is the one that better samples the NIR SED.}.
Finally, we emphasize that this is valid, in general, for undisturbed, face-on nearby Seyfert galaxies without dust lanes, 
and using high angular resolution IR data, dominated by torus emission.

\section{Conclusions}

We have compiled subarcsecond NIR and MIR photometry (1--18 \micron) and MIR spectroscopy (8--13 \micron) for a small 
sample of nearby, undisturbed, nearly face-on Seyfert galaxies without prominent nuclear dust lanes. 
We used average and individual Sy1 and Sy2 SEDs (photometry and spectroscopy) 
to 1) study the sensitivity of different IR wavelengths to the clumpy torus model parameters of Nenkova, 
and 2) derive the minimum combination of IR data needed to constrain torus geometry and intrinsic properties.  
Our major conclusions are as follows:

\begin{itemize}

\item Subarcsecond resolution 8--13 \micron~spectroscopy alone reliably constrains the number of clumps and their
optical depth (N$_0$ and $\tau_V$). Flat silicate features translate in large values of N$_0$ ($\sim$10--15) and 
$\tau_V$ ($\sim$100-150) in the case of the Sy1 galaxies analysed here. On the other hand, shallow absorption features 
are reproduced with large values of N$_0$ ($\sim$8--15) and intermediate $\tau_V$ ($\sim$50) in the case of 
Sy2 galaxies. 

\item It is possible to constrain torus width and inclination ($\sigma$ and $i$) with 
subarcsecond resolution NIR and MIR photometry only, with
steeper IR slopes generally indicating intermediate-to-edge-on views and viceversa.

\item The radial density profile of the clouds $q$ requires fitting of NIR and MIR photometry to be constrained.  
In the case of the Sy1s, adding L- and/or M-band photometry to the MIR data is enough, 
as there is a tight correlation between the MIR slope (measured from $\sim$7 to 15 \micron~by \citealt{Honig10b}) and 
$q$, with flatter slopes producing redder MIR colors. 


\item In addition to NIR photometry, we need 8--13 \micron~spectroscopy and a flat radial profile (i.e. low values of $q$) 
to put realistic constraints on the torus extent ($Y$). However, to have a reliable estimate of the torus size, FIR 
data is required for probing the coolest dust within the torus. For steep radial profiles (large values of $q$: 
the majority of clouds located close to the inner edge of the torus) the NIR/MIR SED is never sensitive to the outer torus extent. 

\item In the case of nearby, undisturbed, nearly face-on Seyferts without dust lanes, the
minimum combination of IR data necessary to reliably constrain all the torus parameters, independently of the 
SED shape, is J+K+M-band photometry + N-band spectroscopy.

\item Despite the fact that it is probing the bulk of the torus emission, the Q-band data analised here (at 18 \micron) lack of 
constraining power when used in combination with 8--13 \micron~spectroscopy, as the latter efficiently reduces the parameters space. 

\end{itemize}




\section*{Acknowledgments}
We are grateful to Sebastian H\"onig for making his ground-based MIR spectrum of NGC\,3227 available to us, and to
Preben Grosb\o l for kindly providing the fully reduced SOFI/Ks intensity map of the galaxy NGC\,5643.
This research was supported by a Marie Curie Intra European Fellowship
within the 7th European Community Framework Programme (PIEF-GA-2012-327934).
CRA ackowledges financial support from the Instituto de Astrof\' isica de Canarias and the 
Spanish Ministry of Science and Innovation (MICINN) through project 
PN AYA2010-21887-C04.04 (Estallidos). CRA is grateful to the Instituto de Astrof\' isica de Cantabria (IFCA) for its hospitality 
during part of this project. 
AAH acknowledges support from the Spanish Plan Nacional de Astronom\' ia y Astrof\' isica under grant
AYA2009-05705-E and from the Augusto G. Linares Program through the Universidad de Cantabria.
OGM and JMRE acknowledge support from the Spanish MICINN through the grant AYA2012-39168-C03-01. 
AAR acknowledges financial support by the Spanish Ministry of Economy and Competitiveness
through projects AYA2010--18029 (Solar Magnetism and Astrophysical Spectropolarimetry) and Consolider-Ingenio 2010 CSD2009-00038. 
AAR also acknowledges financial support through the Ram\'on y Cajal fellowships. CP acknowledges support from NSF-0904421.
This research has made use of the NASA/IPAC Extragalactic Database (NED) which is operated by the Jet Propulsion Laboratory,
California Institute of Technology, under contract with the National Aeronautics and Space Administration. 
Based on observations obtained at the Gemini Observatory, which is operated by the Association of Universities for Research in
Astronomy, Inc., under a cooperative agreement with the NSF on behalf of the Gemini partnership: the National
Science Foundation (United States), the National Research Council (Canada), CONICYT (Chile), the Australian Research
Council (Australia), Minist\'erio da Ci\^encia, Tecnologia e Inova\c{c}$\tilde{a}$o (Brazil) and Ministerio de Ciencia, Tecnolog\' ia
e Innovaci\'on Productiva (Argentina). The Gemini program under which the new data of NGC\,5643 were obtained is GS-2012A-Q-43.
We finally acknowledge useful comments from the anonymous referee.

\appendix

\section{Individual fits.}
\label{appendixA}

\begin{figure*}
\centering
\includegraphics[width=17cm]{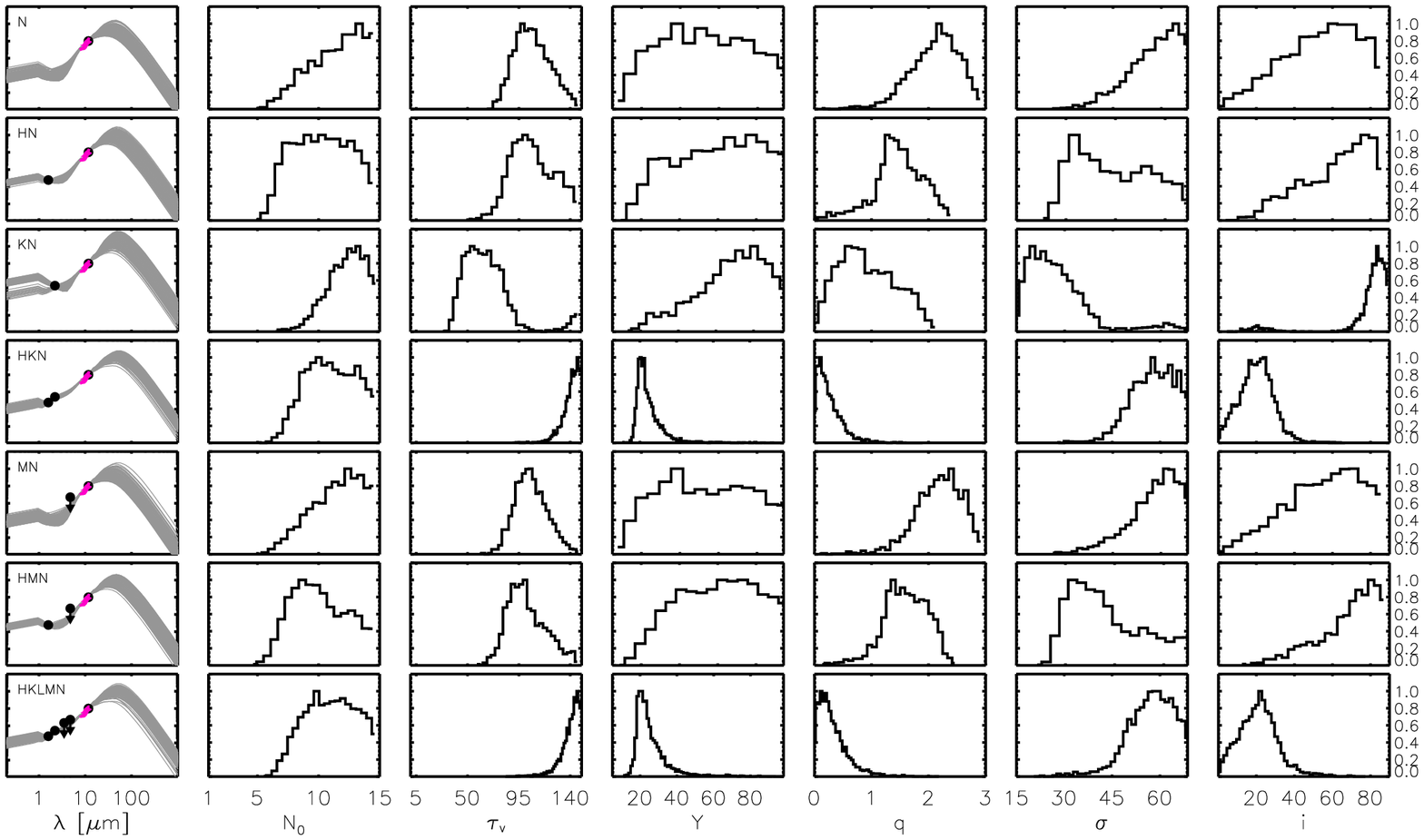}
\caption{Same as in Figure \ref{sy1red}, but for the Sy1.5 NGC\,3227 and trying different combinations of filters. 
\label{ngc3227}}
\end{figure*}

\begin{figure*}
\centering
\includegraphics[width=17cm]{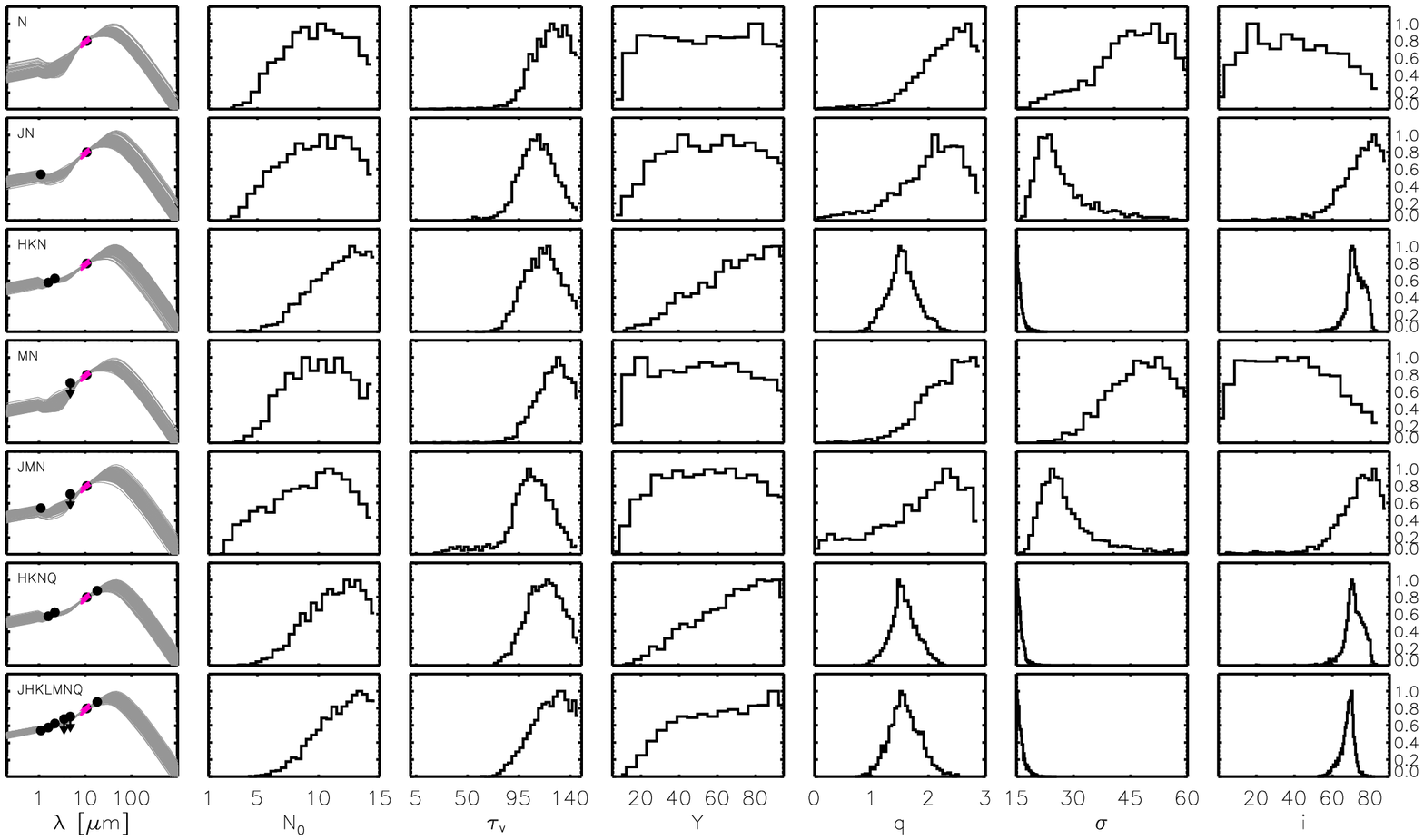}
\caption{Same as in Figure \ref{sy1red}, but for the Sy1.5 NGC\,4151 and trying different combinations of filters. 
\label{ngc4151}}
\end{figure*}

\begin{figure*}
\centering
\includegraphics[width=17cm]{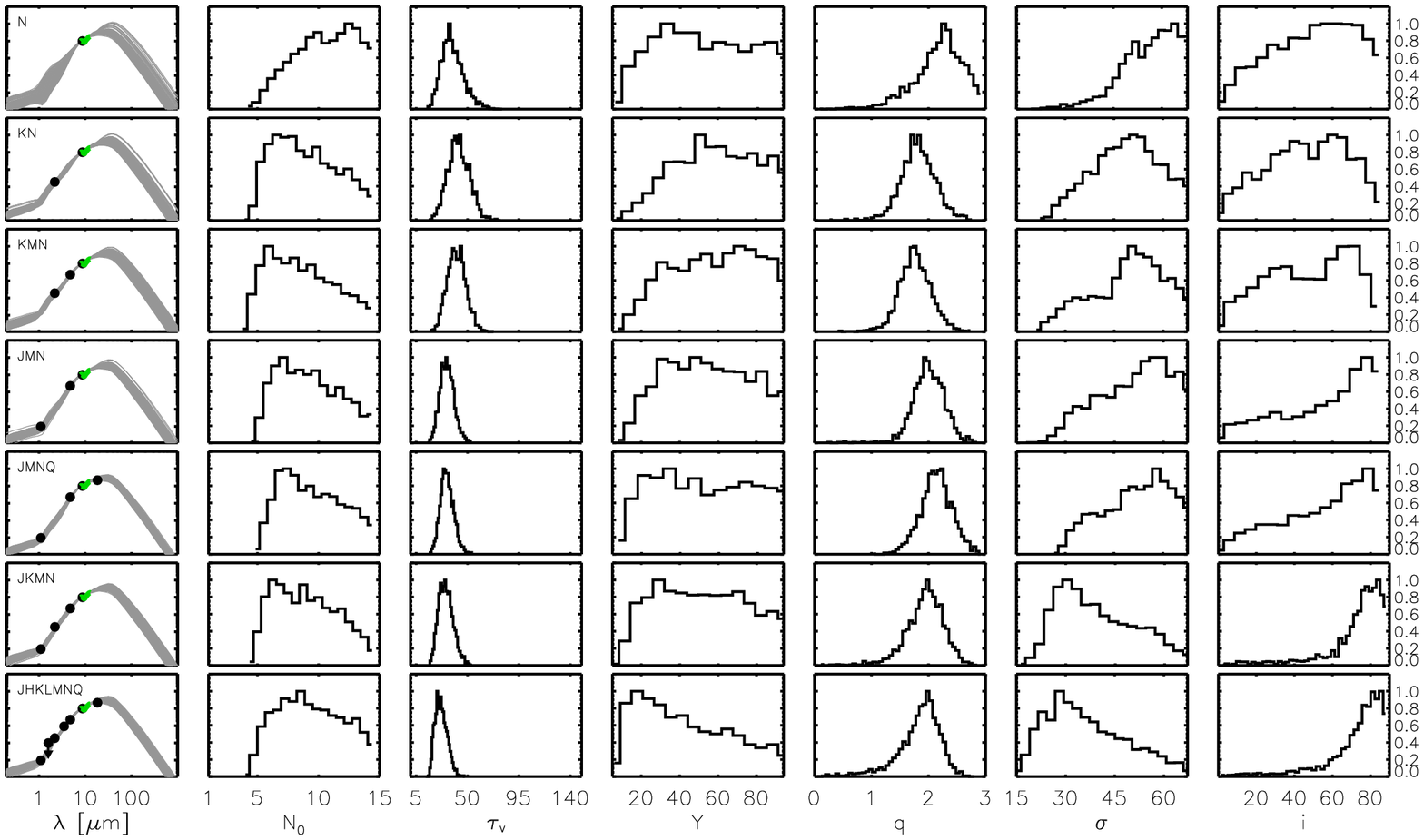}
\caption{Same as in Figure \ref{sy2red}, but for the Sy2 NGC\,1068 and trying different combinations of filters. 
\label{ngc1068}}
\end{figure*}

\begin{figure*}
\centering
\includegraphics[width=17cm]{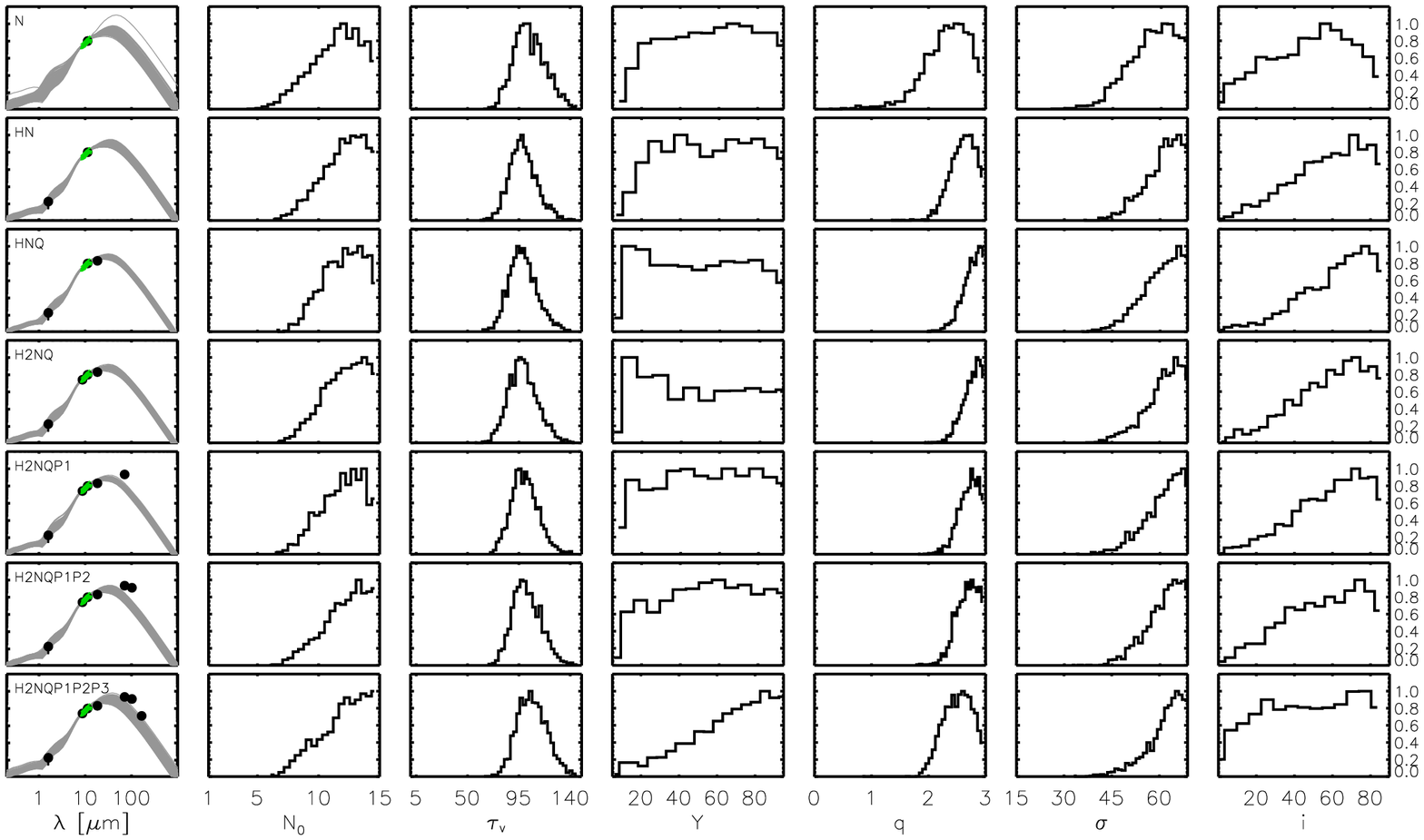}
\caption{Same as in Figure \ref{sy2red}, but for the Sy2 NGC\,3081 and trying different combinations of filters. 
In the last three rows, we have included nuclear FIR fluxes from the 
{\it Herschel Space Observatory} \citep{Ramos11b}, labelled as P1, P2 and P3 (70, 100 and 160 \micron~respectively).
\label{ngc3081}}
\end{figure*}

The results of this work are based on the fits of both the average and individual Sy1 and Sy2 SEDs, 
as different SED shapes are sensitive to different torus properties. 
The individual fits are shown in Figures \ref{ngc3227}, \ref{ngc4151}, \ref{ngc1068} and \ref{ngc3081}. 

\label{lastpage}

\end{document}